\documentclass[%
 reprint,
 superscriptaddress,
 showpacs,preprintnumbers,
 amsmath,amssymb,
 aps,
 prl,
 longbibliography,
]{revtex4-1}

\usepackage{graphicx}
\usepackage{dcolumn}
\usepackage{bm}
\usepackage{color}
\usepackage{ulem}

\begin{document}

\preprint{APS/123-QED}

\title{
Quantum Anomalous Hall Effect in Kagome Ice
}

\author{Hiroaki Ishizuka}
\affiliation{
Department of Applied Physics, University of Tokyo, Tokyo 113-8656, Japan
}

\author{Yukitoshi Motome}%
\affiliation{
Department of Applied Physics, University of Tokyo, Tokyo 113-8656, Japan
}

\date{\today}

\begin{abstract}
An anomalous Hall insulator without magnetic long-range ordering is theoretically reported in the absence of the relativistic spin-orbit coupling. 
It is realized in itinerant electrons coupled with the Ising spins on a $\langle 111 \rangle$ kagome plane of pyrochlore spin ice in applied magnetic field. 
We find that the kagome-ice type local spin correlation in the magnetization plateau state opens a charge gap without magnetic ordering, which results in quantization of the Hall conductivity. 
By Monte Carlo simulation, we identify the anomalous Hall insulating region in the magnetic phase diagram, in addition to another anomalous Hall insulator in a fully-saturated state. 
\end{abstract}


\maketitle

Competing interactions in geometrically frustrated magnets often give rise to peculiar local spin textures or objects composed of several spins. 
A representative example is the two-in two-out spin configuration in each tetrahedron in spin-ice compounds~\cite{Harris1997,Ramirez1999}. 
The local textures or objects have been a matter of intense studies as they are the source of interesting properties in frustrated magnets, such as macroscopic degeneracy with residual entropy and a characteristic power-law spin correlation~\cite{Isakov2004,Henley2005}.

Recently, metallic pyrochlore oxides have opened up a new aspect in the study of geometrically frustrated magnets.
In these compounds, localized spin-ice moments on a pyrochlore lattice interact with itinerant electrons; 
hence, it is expected that the local spin objects affect the electronic and transport properties.
Indeed, various peculiar transport properties were reported, such as unconventional anomalous Hall effects in Nd$_2$Mo$_2$O$_7$~\cite{Taguchi2001} and Pr$_2$Ir$_2$O$_7$~\cite{Machida2007,Machida2009,Balicas2011}, and resistivity minimum in Pr$_2$Ir$_2$O$_7$~\cite{Nakatsuji2006} and Nd$_2$Ir$_2$O$_7$~\cite{Sakata2011}.
Effects of local spin objects on these phenomena were also studied theoretically~\cite{Onoda2003,Udagawa2012}.

The interest in the interplay between local objects and itinerant electrons has spanned to a wider range of systems. 
For instance, extended Falicov-Kimball models on frustrated lattices show non-Fermi liquid behavior~\cite{Udagawa2010} and peculiar metal-insulator transition~\cite{Ishizuka2011}.
Another related study was done on a frustrated double-exchange (DE) model, focusing on the nature of loops emergent from local spin textures~\cite{Jaubert2012}. 
These results indicate that such interplay offers a fertile ground for exploring unconventional electronic and transport properties.

\begin{figure}
   \includegraphics[width=0.9\linewidth]{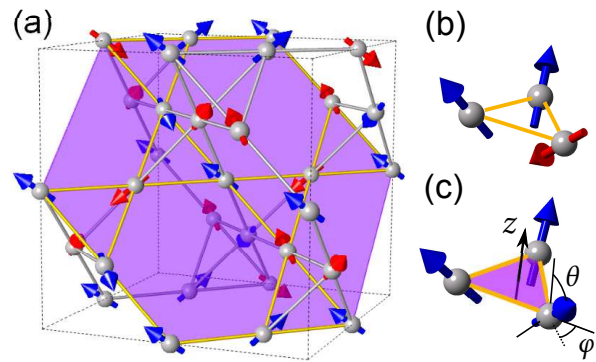}
   \caption{(color online).
   (a) Schematic picture of spin ice on a pyrochlore lattice.
   The shaded plane indicates a $\langle111\rangle$ kagome plane.
   (b) One-in two-out and (c) all-out type spin configuration.
   }
   \label{fig:model}
\end{figure}

In this Letter, we propose that a peculiar spin texture gives rise to an anomalous Hall insulator (AHI) without a magnetic long-range order. 
We show that the so-called kagome ice, which appears in a $\langle 111 \rangle$ kagome plane of the pyrochlore spin ice in applied magnetic field, affects the electronic state significantly by opening a charge gap. 
We numerically show that the insulating kagome ice state exhibits a nonzero quantization of the Hall conductivity induced by the spin scalar chirality, indicating that it is an AHI characterized by the first Chern number.  
Increasing the magnetic field, we find that the AHI state turns into another AHI with saturation of the magnetization; this can be viewed as a topological change between the two insulators. 
The conventional quantum Hall state is associated with the formation of quantized Landau levels in external magnetic field.
Recent studies on topological states of matter revealed that a quantum (spin) anomalous Hall effect occurs in the presence of the relativistic spin-orbit coupling~\cite{Haldane1988,Kane2005,Qi2006,Moore2007,Fu2007}.
Besides, a noncoplanar magnetic order can also give rise to an AHI~\cite{Ohgushi2000,Martin2008}.
Our present result offers yet another AHI without Landau levels, spin-orbit coupling, nor magnetic ordering.
 
We consider spin-ice type Ising moments on a kagome lattice.
The localized Ising moments correspond to those on a $\langle 111 \rangle$ kagome plane of the spin ice model, as shown in Fig.~\ref{fig:model}(a); the anisotropy axis of the Ising spin is along the direction connecting the centers of neighboring two tetrahedra.
When the Ising spins interact with each other by the nearest-neighbor (NN) ferromagnetic interaction, the ground state remains disordered with macroscopic degeneracy, in which all the triangles obey one-in two-out or two-in one-out spin configurations [Fig.~\ref{fig:model}(b)].
An external magnetic field perpendicular to the plane partially lifts the degeneracy by enforcing the upward (downward) triangles to be two-in one-out (one-in two-out), but the ground state is still disordered. 
This is called the kagome ice, which is realized in the magnetization plateau state in spin ice~\cite{Matsuhira2002,Higashinaka2003,Hiroi2003}. 
For a higher field, all-in(out) upward (downward) triangles (magnetic monopoles in the pyrochlore spin ice~\cite{Castelnovo2008}) are introduced into the kagome ice, resulting in a first-order transition to a saturated state.

To clarify the effects of such characteristic local spin textures on itinerant electrons, we here consider a single-band DE model~\cite{Zener1951,Anderson1955} with the Ising spins on a kagome lattice;
\begin{eqnarray}
H = - \sum_{\langle i,j \rangle} ( t_{ij} c^\dagger_{i} c_{j} + \text{H.c.} ) - \sum_i h_z \cos \theta_i.
\label{eq:H}
\end{eqnarray}
Here, $c_i$ ($c_i^\dagger$) is the annihilation (creation) operator of an itinerant electron at $i$th site, whose spin index is dropped as the spin is completely aligned parallel to the localized spin $\mathbf{S}_i$ at each site.  
The anisotropy axis of the localized spin depends on the sublattice; ${\bf S}_i=(S_i^x,S_i^y,S_i^z)=S(\sin\theta_i\cos\varphi_i,\sin\theta_i\sin\varphi_i,\cos\theta_i)$, where  $(\theta_i, \varphi_i) = (\arccos(\pm\frac13), \frac{2\pi}{3}n_{\rm s} \pm \frac{\pi}{2})$ for the sublattice $n_{\rm s}=1,2,3$ and $S=1$ [see Fig.~\ref{fig:model}(c)].
The effective transfer integral $t_{ij}$ depends on the relative angle of neighboring Ising spins, which is given by $t_{ij} = t(\cos\frac{\theta_i}{2}\cos\frac{\theta_j}{2}+ \sin\frac{\theta_i}{2}\sin\frac{\theta_j}{2}e^{-{\rm i}(\varphi_i-\varphi_j)})$.
The sum $\langle i,j \rangle$ is taken over NN sites on the kagome lattice. 
In Eq.~(\ref{eq:H}), the effect of external magnetic field is taken into account by the Zeeman term only for Ising spins for simplicity.
Hereafter, we set the energy unit $t=1$, the length of Bravais lattice vector $a = 1$, the Boltzmann constant $k_{\rm B} = 1$, and the unit of conductance $e^2/h = 1$ ($e$ is the elementary charge and $h$ is the Planck constant).

In this model, the spin-charge coupling induces an effective ferromagnetic interaction between the localized spins at general filling~\cite{Zener1951,Anderson1955}.
Although the NN ferromagnetic interaction favors macroscopic degeneracy as described above, the spin-charge coupling also gives rise to further-neighbor interactions, which generally lift the degeneracy and select a magnetic order in the ground state.
However, as the further-neighbor interactions are usually much weaker than the NN one, the liquid-like state with strong local correlations of two-in one-out or one-in two-out types is expected to emerge in the intermediate temperature ($T$) region between a high-$T$ disordered state and the low-$T$ ordered phase.
This is indeed confirmed by the finite-$T$ Monte Carlo (MC) simulation as we discuss later. 
In the following, we focus on how the electronic and transport properties in the liquid-like state evolve in applied magnetic field.

First, for simplicity, we study the problem by taking a simple average (arithmetic mean) over different configurations of spins.
Instead of the magnetic field $h_z$, we control the net magnetization along the $z$ direction per triangle, $m_z = \frac{3}{N} \sum_i \cos \theta_i$, where $N$ is the number of sites. 
The average is taken over different spin configurations with a fixed ratio of different types of triangles as follows. 
For $0 \leq m_z \leq 1/3$, we consider the manifold in which all the triangles are of two-in one-out or one-in two-out type; their ratio is controlled so as to realize the given value of $m_z$. 
At $m_z=1/3$, all the upward (downward) triangles are in the two-in one-out (one-in two-out) configuration, which is the kagome ice. 
For $m_z>1/3$, we introduce all-in upward and all-out downward triangles.
Eventually, at the saturation to $m_z=1$, the system shows a long-range order of alternating all-in and all-out triangles. 
At each value of $m_z < 1$, we calculate the electronic state of the model in Eq.~(\ref{eq:H}) by numerical diagonalization with taking the simple average over different spin configurations generated by a loop update method~\cite{Ishizuka2011}. 
The conductivity is also numerically obtained by using the Kubo formula~\cite{note_numerics}. 

\begin{figure}
   \includegraphics[width=0.8\linewidth]{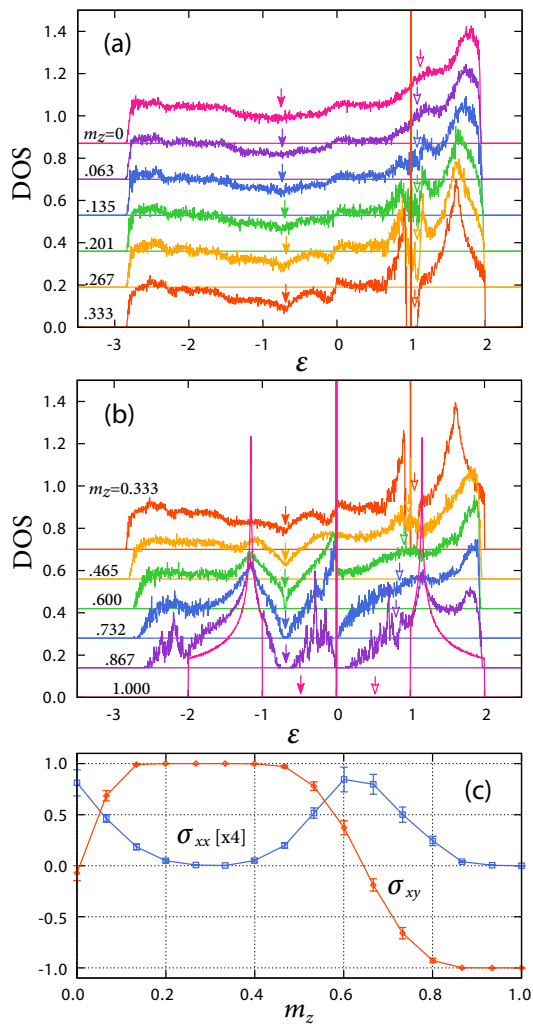}
   \caption{(color online).
   (a), (b) DOS for different $m_z$ calculated by the simple average. See the text for details.
   Solid and open arrows indicate the Fermi levels for $n=1/3$ and $n=2/3$, respectively.
   The data for different $m_z$ are plotted with the offset of $0.14$ and $0.19$ in (a) and (b), respectively.
   (c) shows the longitudinal and transverse conductivities at $n=2/3$ as a function of $m_z$.
   }
   \label{fig:dos}
\end{figure}

Figures~\ref{fig:dos}(a) and \ref{fig:dos}(b) show the electronic density of states (DOS) for different values of $m_z$.
The solid and open arrows on the spectra indicate the Fermi levels for the electron filling $n=1/3$ and $n=2/3$, respectively ($n = \frac1N \sum_i \langle c_i^\dagger c_i \rangle$).
Figure~\ref{fig:dos}(a) shows the results for $0 \leq m_z \leq 1/3$, where all the triangles are in two-in one-out or one-in two-out configurations.
The result indicates that, although there is no long-range magnetic order, the characteristic spatial correlations in the kagome ice manifold develops a charge gap at $n = 2/3$ near $m_z=1/3$~\cite{note_flatband}.
Figure~\ref{fig:dos}(b) shows the results for $1/3 \leq m_z \leq 1$, where the all-in upward or all-out downward triangles (magnetic monopoles) are introduced into the kagome ice manifold.
At the Fermi level for $n=1/3$, the introduction of monopoles leads to a dip in the spectra, and opens a full gap for $m_z\gtrsim 0.7$.
On the other hand, the energy gap at $n=2/3$ and $m_z=1/3$ is closed by introducing monopoles.
However, for $m_z \gtrsim 0.7$, DOS develops a dip again and shows an energy gap at $m_z \simeq 1$~\cite{note_m_z=1}.

To gain further insight into the peculiar changes, we calculated the longitudinal ($\sigma_{xx}$) and transverse ($\sigma_{xy}$) conductivities.
Figure~\ref{fig:dos}(c) shows the results at $n=2/3$.
For $0 \leq m_z \leq 1/3$, $\sigma_{xx}$ shows monotonic decrease in accordance with the growth of energy gap in DOS.
On the other hand, $\sigma_{xy}$ shows monotonic increase while increasing $m_z$, and surprisingly, the numerical results show a quantized value at $\sigma_{xy}=1$ in the gapped state near $m_z=1/3$~\cite{notes_error}.
For $m_z>1/3$, $\sigma_{xx}$ increases and shows a hump at $m_z \sim 2/3$; finally it decreases to zero in the all-in/all-out insulator at $m_z=1$. 
Correspondingly, $\sigma_{xy}$ decreases from $1$ with showing a sign change at $m_z \sim 2/3$, and converges to another quantized value $\sigma_{xy}=-1$ at $m_z=1$.

The non-monotonic change of the Hall conductivity $\sigma_{xy}$ is explained by the Berry phase mechanism~\cite{Ohgushi2000,Loss1992,Ye1999}. 
In the previous study by assuming a ${\bf q}=0$ magnetic order~\cite{Ohgushi2000}, it was shown that the Hall conductance depends on the scalar chirality of localized spins in the three-site unit cell, ${\bf S}_i\cdot {\bf S}_j\times{\bf S}_k$.
A similar mechanism was shown to work through the fluctuations~\cite{Onoda2003}.
Our result of $\sigma_{xy}$ is understood by considering that two-in one-out (one-in two-out) upward (downward) triangles bring a negative chirality $-\frac{4}{3\sqrt{3}}$ each and that magnetic monopoles bring a positive chirality $+\frac{4}{3\sqrt{3}}$ each.

The remarkable point in our results is the quantization of $\sigma_{xy}$ at $+1$ in the gapped kagome ice state and its switching to $-1$ accompanied by the closing and reopening of the energy gap at $n=2/3$ and associated hump in $\sigma_{xx}$. 
The change in $\sigma_{xy}$ as well as the hump in $\sigma_{xx}$ suggests a transition between the kagome ice at $m_z=1/3$ and all-in/all-out ordered state at $m_z=1$. 
The latter ordered state was shown to be an AHI characterized by the  first Chern number $C=-1$~\cite{Ohgushi2000}. 
On the other hand, the kagome ice at $m_z=1/3$ is a paramagnetic state with constrained local spin configurations;
hence, the quantization of $\sigma_{xy}$ as well as the gap opening in DOS is highly nontrivial, suggesting that the liquid-like kagome ice is also another AHI.
The closing and reopening of the energy gap as well as the hump of $\sigma_{xx}$ is interpreted as a phase transition between topologically different insulators with a transient metallic state in between.

The above calculations suggest that a similar change between the kagome ice and all-in/all-out ordered state can take place in the model (\ref{eq:H}) in applied magnetic field $h_z$. 
To confirm this, we next investigate the thermodynamic behavior by a MC simulation.
In this method, the MC updates of localized spins are done by directly evaluating the effective action from itinerant electrons using exact diagonalization~\cite{Yunoki1998,note_numerics2}.

\begin{figure}
   \includegraphics[width=\linewidth]{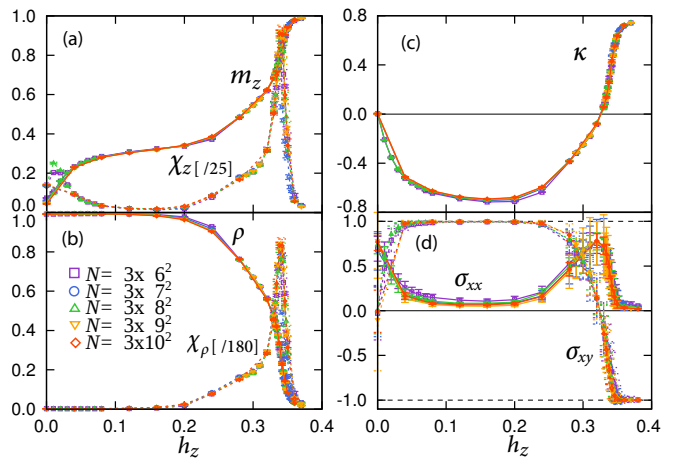}
   \caption{ (color online).
   MC results for (a) $m_z$ and $\chi_z$, (b) $\rho$ and $\chi_{\rho}$ (c) $\kappa$, and (d) $\sigma_{xx}$ and $\sigma_{xy}$ at $n=2/3$ and $T=0.03$.
   The data were calculated for the system sizes ranging from $N=3\times6^2$ to $3\times 10^2$. 
   }
   \label{fig:mcdata}
\end{figure}

Figure~\ref{fig:mcdata} shows the MC results at $T=0.03$ and $n=2/3$ in applied field $h_z$.
All the results consistently indicate two sharp crossovers while increasing $h_z$; 
one is from the zero-field state to the kagome ice state by switching on $h_z$, and the other to the all-in/all-out ordered state at $h_z \sim 0.34$. 
Figure~\ref{fig:mcdata}(a) depicts the results for $m_z$ and its susceptibility $\chi_z$.
They clearly indicate the presence of $1/3$ magnetization plateau for $0.1\lesssim h_z \lesssim 0.3$ followed by rapid increase of $m_z$ for larger $h_z$ and a peak of $\chi_z$ at $h_z=0.341(3)$.
As shown in Fig.~\ref{fig:mcdata}(b), the ratio of two-in one-out and one-in two-out configurations in the system, $\rho$, stays close to $1$ in the plateau regime, while it decreases for larger $h_z$; 
the corresponding susceptibility $\chi_\rho$ shows a sharp peak at $h_z = 0.340(3)$.
Figure~\ref{fig:mcdata}(c) shows the net scalar chirality $\kappa=\frac3{2N}\sum_{(i,j,k)}\langle{\bf S}_i\cdot {\bf S}_j\times{\bf S}_k \rangle$, where the sum is taken over all the triangles and the indices $(i,j,k)$ are in the counterclockwise order in each triangle.
It approaches and stays near the minimum value $-\frac{4}{3\sqrt{3}}$, followed by rapid increase for a higher field with showing a sign change at $h_z = 0.325(5)$. 
Correspondingly, $\sigma_{xy}$ becomes $1.0$ associated with strong suppression of $\sigma_{xx}$ in the kagome ice plateau region, and shows a sign change at $h_z=0.328(7)$ accompanied by a hump of $\sigma_{xx}$.
The result clearly indicates that the second crossover occurs between two topologically different insulating states, the kagome ice and all-in/all-out ordered states.

\begin{figure}
   \includegraphics[width=0.9\linewidth]{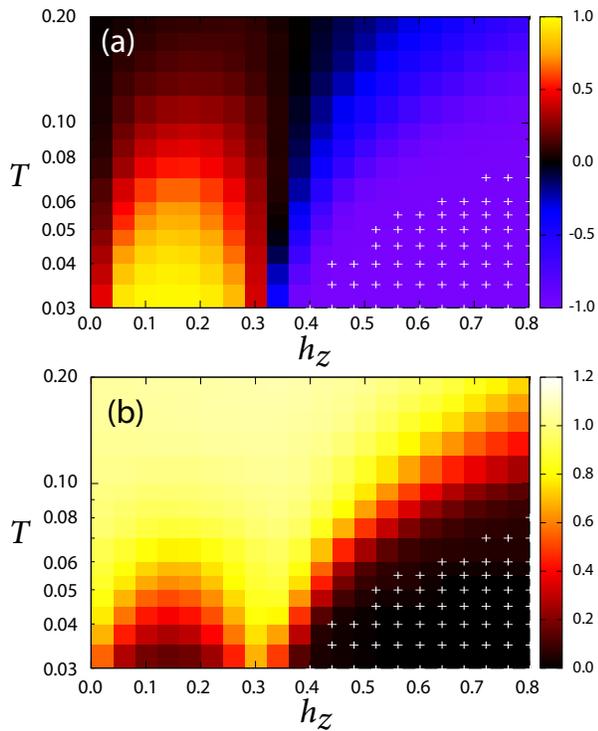}
   \caption{(color online).
   Contour plots of the MC results for (a) $\sigma_{xy}$ and (b) $\sigma_{xx}$ in the $h_z$-$T$ plane. 
   The calculations were done for $N=3\times 6^2$ size systems. 
   The white crosses show the parameters at which the MC acceptance rate becomes lower than $1\%$. 
   }
   \label{fig:hTdiagram}
\end{figure}

The two insulating phases are clearly observed in the contour plots of $\sigma_{xy}$ and $\sigma_{xx}$ in the $h_z$-$T$ plane, as shown in Fig.~\ref{fig:hTdiagram}.
Figure~\ref{fig:hTdiagram}(a) clearly indicates two regimes with a finite Hall conductivity $\sigma_{xy}$ of an opposite sign: a dome-like region centered at $h_z \sim 0.15$, and another one for $h_z \gtrsim 0.34$.
The two regimes are also clearly indicated by strong suppression of $\sigma_{xx}$, as shown in Fig.~\ref{fig:hTdiagram}(b).
Thus, the phase diagram shows that the model in Eq.~(\ref{eq:H}) exhibits two different insulating regimes with different topological characters, as expected in the analysis by using the simple average in Fig.~\ref{fig:dos}. 

Our comprehensive analyses provide convincing evidence of the existence of kagome ice AHI in the intermediate-$T$ liquid-like regime. 
This is a peculiar topological state of matter in the absence of Landau levels, spin-orbit coupling, and magnetic ordering.
Upon further decreasing $T$, it is expected that effective further-neighbor interactions induced by the spin-charge coupling stabilize some magnetic order, whose ordering pattern may become complicated depending on $h_z$ as well as the electron density.
It is left for future study to investigate the full phase diagram including the low-$T$ ordered phases. 
When the model is extended to more realistic situation on the three-dimensional pyrochlore lattice, the AHI behavior might be smeared out, but the non-monotonic change of $\sigma_{xy}$ is likely to remain in applied magnetic field.
It is interesting to note that such non-monotonic behavior of the Hall resistivity was observed in the $\langle 111 \rangle$ magnetic field in Pr$_2$Ir$_2$O$_7$~\cite{Machida2007,Balicas2011}.

The authors thank A. Shitade and M. Udagawa for fruitful discussions and T. Kitagawa for helpful comments. 
H.I. is supported by Grant-in-Aid for JSPS Fellows.
This research was supported by KAKENHI (No.19052008, 21340090, 22540372, and 24340076), Global COE Program ``the Physical Sciences Frontier", the Strategic Programs for Innovative Research (SPIRE), MEXT, and the Computational Materials Science Initiative (CMSI), Japan.


\begin{thebibliography}{99}
\bibitem{Harris1997}        M. J. Harris, S. T. Bramwell, D. F. McMorrow, T. Zeiske, and K. W. Godfrey, Phys. Rev. Lett. {\bf 79}, 2554 (1997).
\bibitem{Ramirez1999}       A. P. Ramirez, A. Hayashi, R. J. Cava, R. Siddharthan, and B. S. Shastry, Nature (London) {\bf 399}, 333 (1999).
\bibitem{Isakov2004}        S. V. Isakov, K. Gregor, R. Moessner, and S. L. Sondhi, Phys. Rev. Lett. {\bf 93}, 167204 (2004).
\bibitem{Henley2005}        C. L. Henley, Phys. Rev. B {\bf 71}, 014424 (2005).
\bibitem{Taguchi2001}       Y. Taguchi, Y. Oohara, H. Yoshizawa, N. Nagaosa, and Y. Tokura, Science {\bf 291}, 2573 (2001).
\bibitem{Machida2007}       Y. Machida, S. Nakatsuji, Y. Maeno, T. Tayama, T. Sakakibara, and S. Onoda, Phys. Rev. Lett. {\bf 98}, 057203 (2007).
\bibitem{Machida2009}       Y. Machida, S. Nakatsuji, S. Onoda, T. Tayama, and T. Sakakibara, Nature (London) {\bf 463}, 210 (2009).
\bibitem{Balicas2011}       L. Balicas, S. Nakatsuji, Y. Machida, and S. Onoda, Phys. Rev. Lett. {\bf 106}, 217204 (2011).
\bibitem{Nakatsuji2006}     S. Nakatsuji, Y. Machida, Y. Maeno, T. Tayama, T. Sakakibara, J. van Duijn, L. Balicas, J. N. Millican, R. T. Macaluso, and J. Y. Chan, Phys. Rev. Lett. {\bf 96}, 087204 (2006).
\bibitem{Sakata2011}        M. Sakata, T. Kageyama, K. Shimizu, K. Matsuhira, S. Takagi, M. Wakeshima, and Y. Hinatsu, Phys. Rev. B {\bf 83}, 041102(R) (2011).
\bibitem{Onoda2003}         S. Onoda and N. Nagaosa, Phys. Rev. Lett. {\bf 90}, 196602 (2003).
\bibitem{Udagawa2012}       M. Udagawa, H. Ishizuka, and Y. Motome, Phys. Rev. Lett. {\bf 108}, 066406 (2012).
\bibitem{Udagawa2010}       M. Udagawa, H. Ishizuka, and Y. Motome, Phys. Rev. Lett. {\bf 104}, 226405 (2010).
\bibitem{Ishizuka2011}      H. Ishizuka, M. Udagawa, and Y. Motome, Phys. Rev. B {\bf 83}, 125101 (2011).
\bibitem{Jaubert2012}       L. D. C. Jaubert, S. Piatecki, M. Haque, and R. Moessner, Phys. Rev. B {\bf 85}, 054425 (2012).
\bibitem{Haldane1988}       F. D. M. Haldane, Phys. Rev. Lett. {\bf 61}, 2015 (1988).
\bibitem{Kane2005}          C. L. Kane and E. J. Mele, Phys. Rev. Lett. {\bf 95}, 146802 (2005).
\bibitem{Qi2006}            X. L. Qi, Y. S. Wu, and S. C. Zhang, Phys. Rev. B {\bf 74}, 085308 (2006).
\bibitem{Moore2007}         J. E. Moore and L. Balents, Phys. Rev. B {\bf 75}, 121306(R) (2007).
\bibitem{Fu2007}            L. Fu, C. L. Kane, and E. J. Mele, Phys. Rev. Lett. {\bf 98}, 106803 (2007).
\bibitem{Ohgushi2000}       K. Ohgushi, S. Murakami, and N. Nagaosa, Phys. Rev. B {\bf 62}, 6065(R) (2000).
\bibitem{Martin2008}        I. Martin and C. D. Batista, Phys. Rev. Lett. {\bf 101}, 156402 (2008).
\bibitem{Matsuhira2002}	    K. Matsuhira, Z. Hiroi, T. Tayama, S. Takagi, and T. Sakakibara, J. Phys.: Cond. Matter {\bf14}, L559 (2002).
\bibitem{Higashinaka2003}   R. Higashinaka, H. Fukazawa, and Y. Maeno, Phys. Rev. B {\bf 68}, 014415 (2003).
\bibitem{Hiroi2003}         Z. Hiroi, K. Matsuhira, S. Takagi, T. Tayama, and T. Sakakibara, J. Phys. Soc. Jpn. {\bf 72}, 411 (2003).
\bibitem{Castelnovo2008}    C. Castelnovo, R. Moessner, and S. L. Sondhi, Nature {\bf 451}, 42 (2008).
\bibitem{Zener1951}         C. Zener, Phys. Rev. {\bf 82}, 403 (1951).
\bibitem{Anderson1955}      P. W. Anderson and H. Hasegawa, Phys. Rev. {\bf 100}, 675 (1955).
\bibitem{note_numerics}		The calculations were done on $4^2$ superlattices of $N=3\times 18^2$ sites and $T=10^{-5}$ by averaging over 32 realizations of different spin configurations. Conductivities are calculated with assuming a small inelastic scattering rate $\tau^{-1} = (3\times 18^2\times 4^2)^{-1}$.
\bibitem{note_flatband}     The flat band at $\varepsilon=1$ originates from the bound states within the hexagons with all spins ``in" or all spins ``out", as a consequence of the quantum phase interference.
\bibitem{note_m_z=1}		In the all-in/all-out ordered state at $m_z=1$, DOS has a particle-hole symmetric form with two gaps at $n=1/3$ and $n=2/3$ above and below the central flat band~\cite{Ohgushi2000}.
\bibitem{notes_error}       $\sigma_{xy}$ becomes $+1$ for all kagome ice configurations that we generated numerically within errors of the order of $10^{-6}$. Small deviations presumably come from a small $T$ introduced in the calculations.
\bibitem{Loss1992}          D. Loss and P. M. Goldbart, Phys. Rev. B {\bf 45}, 13544 (1992).
\bibitem{Ye1999}            J. Ye, Y. B. Kim, A. J. Millis, B. I. Shraiman, P. Majumdar, and Z. Te$\rm\check{s}$anovic, Phys. Rev. Lett. {\bf 83}, 3737 (1999).
\bibitem{Yunoki1998}        S. Yunoki, J. Hu, A. L. Malvezzi, A. Moreo, N. Furukawa, and E. Dagotto, Phys. Rev. Lett. {\bf 80}, 845 (1998).
\bibitem{note_numerics2}	MC calculations were conducted up to $N=3\times10^2$ with typically 20000 MC measurements after 6000 steps of thermalization. Conductivities were calculated by taking $\tau=0.05$.
\end{thebibliography}
\end{document}